# Novel high-pressure calcium carbonates

Xi Yao[1,2], Congwei Xie[1,5*], Xiao Dong[3,4*], Artem R. Oganov[5,1,6*] and Qingfeng Zeng[2]

[1]*International Center for Materials Discovery, School of Materials Science and Engineering, Northwestern Polytechnical University, Xi'an, Shaanxi 710072, PR China*

[2]*Science and Technology on Thermostructural Composite Materials Laboratory, School of Materials Science and Engineering, Northwestern Polytechnical University, Xi'an, Shaanxi 710072, PR China*

[3] *School of Physics, Nankai University, Tianjin 300071, China.*

[4] *Center for High Pressure Science and Technology Advanced Research, Beijing 100193, China*

[5]*Skolkovo Institute of Science and Technology, 3 Nobel Street, Moscow 143026, Russia*

[6]*Moscow Institute of Physics and Technology, 9 Institutskiy Lane, Dolgoprudny City, Moscow Region 141700, Russia*

## Abstract

Calcium and magnesium carbonates are believed to be the host compounds for most of the oxidized carbon in the Earth's mantle. Here, using evolutionary crystal structure prediction method USPEX, we systematically explore the $MgO$-$CO_2$ and $CaO$-$CO_2$ systems at pressures ranging from 0 to 160 GPa to search for thermodynamically stable magnesium and calcium carbonates. While $MgCO_3$ is the only stable magnesium carbonate, three calcium carbonates are stable under pressure: well-known $CaCO_3$, and newly predicted $Ca_3CO_5$ and $CaC_2O_5$. $Ca_3CO_5$ polymorphs are found to contain isolated orthocarbonate $CO_4^{4-}$ tetrahedra, and are stable at relatively low pressures (>11 GPa), whereas $CaC_2O_5$ is stable above 33 GPa and its polymorphs feature polymeric motifs made of $CO_4$-tetrahedra. Detailed analysis of chemical stability of $CaCO_3$, $Ca_3CO_5$ and $CaC_2O_5$ in the environment typical of the Earth's lower mantle reveals that none of these compounds can exist in the Earth's lower mantle. We conclude that $MgCO_3$ is the main host of oxidized carbon throughout the lower mantle.

## Introduction

Behaviour of carbon in the Earth's mantle is important for the global carbon cycle. The generally accepted view is that mantle carbon exists in a number of reduced, neutral, and oxidized forms (i.e., $Fe_3C$ cementite, diamond, carbonates) [1].

Over the past few decades, magnesium and calcium carbonates ($MgCO_3$ and $CaCO_3$) have received considerable attention because they are believed to be the host compounds for most of the oxidized carbon in the mantle [2-6]. Zero-temperature calculations of Pickard and Needs predicted that $CaCO_3$ will become more favorable than $MgCO_3$ at pressures above 100 GPa at mantle chemistry and therefore should be present in the lowermost matle [6], but thermal effects could change this conclusion (and this is indeed the case). Moreover, all previous previous works assumed that compositions known at atmospheric pressure ($CaCO_3$, $MgCO_3$) as the only possibilities. Recent works [7-9] proved that chemistry is greatly altered by pressure: new unexpected compound appear so often that they are more of a rule than exception. It is, therefore, necessary to check for additional possible carbonates.

At mantle pressures, a series of phase transitions occur in $MgCO_3$ and $CaCO_3$. Tables S1 and S2 list high-pressure forms of $MgCO_3$ and $CaCO_3$ predicted in previous works [3,5,6]. From zero pressure up to the pressure of the core-mantle boundary (136 GPa), both $MgCO_3$ and $CaCO_3$ will experience several interesting phase transitions. For example, it was predicted that polymorphs of $CaCO_3$ stable below 76 GPa [6] (or 75 GPa according to our calculations) feature $CO_3$-triangles, while chains of $CO_4$ tetrahedra are present in the higher-pressure form of $CaCO_3$.

At ambient pressure and temperature, carbon atom in $CO_2$ has *sp*-hybridization with linear geometry and 2-fold coordination, while in $CO_3^{2-}$ it has $sp^2$-hybridization resulting in planar triangular geometry and 3-fold coordination. $sp^3$-hybridization (resulting in $CO_4$ tetrahedra) is unfavorable due to very small size of $C^{4+}$ cation compared to $O^{2-}$ atom: at realistic (very short) C-O distances, steric O-O repulsion would be too high. However, at high pressure, carbon prefers to be in the $sp^3$ state and behaves in many ways akin to silicon at normal pressure. High coordination gives volume advantage, which offsets the steric effects.

In this work, we assess the traditional assumption of $CaCO_3$ and $MgCO_3$ stoichiometries of calcium and magnesium carbonates. As will be detailed below, new carbonates are indeed predicted, and we examine their stability at different pressures and temperatures, and in the chemical environment of the lower mantle of the Earth.

# Computational methodology

To search for stable magnesium and calcium carbonates at mantle pressures, we have explored the MgO-$CO_2$, CaO-$CO_2$ Mg-C-O, Ca-C-O and MgO-CaO-$CO_2$ systems using the variable-composition evolutionary algorithm (EA) technique, as implemented in the USPEX code [10-13]. Here we performed EA crystal structure predictions in the pressure range from 0 to 160 GPa with up to 40 atoms in the primitive unit cell. The first generation of structures was created randomly. In all subsequent generations, structures were produced by heredity (40%), symmetric random generator (20%), softmutation (20%) and transmutation (20%) operators, and the best 60% of previous generation were used as parents to generate the next generation of structures. For all structures generated by USPEX, structure relaxations and total energy calculations were performed using the VASP code [14] in the framework of density functional theory [15]. In these calculations, we used the Perdew-Burke-Ernzerhof generalized gradient approximation functional (PBE-GGA [16]) to treat exchange-correlation, and the all-electron projector augmented wave (PAW [17]) method to describe core-valence interactions - $3s^2 3p^6 4s^2$, $3s^2$, $2s^2 2p^2$ and $2s^2 2p^4$ shells were treated as valence for Ca, Mg, C and O, respectively. The plane-wave kinetic energy cutoff of 600 eV and uniform *k*-point meshes for sampling the Brillouin zone with reciprocal-space resolution of $2\pi\times0.05$ Å$^{-1}$ were employed. Once stable compounds and structures were found, their properties were computed with denser *k*-points meshes, which had reciprocal-space resolution of $2\pi\times0.03$ Å$^{-1}$.

# Results and discussions

## 1. Phase stability at mantle pressures

We have performed crystal structure predictions at 0 GPa, 15 GPa, 20 GPa, 40 GPa, 60 GPa, 80 GPa 100 GPa and 160 GPa for the CaO-$CO_2$ system, at 0 GPa, 60 GPa, 100 GPa and 160 GPa for the MgO-$CO_2$ system, and 25 GPa, 50 GPa, 100 GPa, 130 GPa for ternary systems: Mg-C-O, Ca-C-O and MgO-CaO-$CO_2$. At a given pressure, stable compounds were determined by the thermodynamic convex hull construction. As shown in Fig. 1(a), among all possible magnesium carbonates, only $MgCO_3$ was found to be on the convex hull. This indicates that, except $MgCO_3$, no other magnesium carbonates can be thermodynamically stable at mantle pressures. At the same time, in the CaO-$CO_2$ system, besides the well-known $CaCO_3$, we have discovered two hitherto unreported thermodynamically stable calcium carbonates $Ca_3CO_5$ and $CaC_2O_5$ (and one near-ground state compound $Ca_2CO_4$), as shown in Fig. 1(b). Phonon calculations show that the two new stable calcium carbonate phases are dynamically stable, i.e. have no imaginary phonon frequencies, as shown in Fig. 2. Unexpectedly, we also found that three high-pressure forms of $CaC_2O_5$ (*Pc*, *Fdd*2 and *C*2) and *Cmcm*-$Ca_3CO_5$ can maintain dynamical stability at 0 GPa, see Fig. S1 in Supplementary Materials. This means that $Ca_3CO_5$ and $CaC_2O_5$ can be quenchable to ambient conditions at low temperature.

By calculating enthalpy-pressure curves for all stable compounds in the CaO-$CO_2$ and MgO-$CO_2$ systems (see Fig. S2 in Supplementary Materials), we have obtained their pressure-composition phase diagrams at pressures up to 160 GPa (see Fig. 3). As shown in Fig. 3, stable phases and phase transition pressures in MgO, CaO, $CO_2$, $MgCO_3$ and $CaCO_3$ are in good agreement with previous studies [3,5,6, 18-20]. We note that, for $CaCO_3$, Smith *et al*. recently proposed a new $P2_1/c$-II phase which is stable between 27.2 GPa and 37.5 GPa[21]. Unfortunately, we missed this structure in the present structural predictions; moreover, we are unable to include this structure in the present study because of the lack of its detailed structural parameters. Considering the slight energy difference between $P2_1/c$-II phase and $P2_1/c$-l of $CaCO_3$, we should say that $P2_1/c$-II phase of $CaCO_3$ will not affect our main conclusion too much. Phase transformations of $Ca_3CO_5$ and $CaC_2O_5$ are as follows: (1) For $Ca_3CO_5$, the orthorhombic *Cmcm* phase is predicted to become stable at 11 GPa, and to transform to tetragonal *I*4/*mcm* phase at 55 GPa; (2) For $CaC_2O_5$, above 33 GPa, four stable phases (low-pressure *Pc*, *Fdd*2, high-pressure *Pc* and *C*2) are predicted. Structural transformation from low-pressure monoclinic *Pc*-phase to orthorhombic *Fdd*2-phase occurs at 38 GPa, and from *Fdd*2-phase to high-pressure monoclinic *Pc*-phase at 72 GPa, and from high-pressure *Pc*-phase to monoclinic *C*2-phase at 82 GPa. Fig.4 shows the computed equations of state (EOS) of all stable calcium carbonates. One can see that pressure-induced phase transitions in $CaCO_3$ and $Ca_3CO_5$, but not in $CaC_2O_5$ are accompanied by large volume discontinuities.

It should be possible to synthesize the newly predicted calcium carbonates ($Ca_3CO_5$ and $CaC_2O_5$). Several phases of $MgCO_3$ and $CaCO_3$, previously predicted by our method and similar techniques, have already been confirmed by experiment, such as $C2/m$ [5] and $P2_1$ [5] phases in $MgCO_3$, and $P2_1/c$ [6,21], *Pmmn* [3,22] and thermodynamically metastable *P*-1 [3,23] in $CaCO_3$. Considering hundreds of papers where it was assumed that $CaCO_3$ is the only possible calcium carbonate, and the importance of calcium carbonate for fundamental chemistry and physics and the hot ongoing quest for $sp^3$ (tetrahedral) carbonates, we believe that our newly predicted calcium carbonates will stimulate experiments.

## 2. Crystal structures of stable calcium carbonates

Crystal structures of the predicted stable and metastable calcium carbonates, visualized by VESTA package [24], are shown in Fig. 5. At mantle pressures, crystal structures of $CaCO_3$ have been carefully studied before [3,6]. With the increase of pressure, $CaCO_3$ successively adopts five phases (calcite *R*-3*c*, aragonite *Pnma*, low-pressure $P2_1/c$, post-aragonite *Pmmn*, and high-pressure $P2_1/c$). As shown in Fig. 5(a)-(e), we can find that the former four phases contain triangular $CO_3^{2-}$ ions with $sp^2$-hybridization, while the fifth phase adopts pyroxene-type structure with chains of corner-linked $CO_4^{4-}$ tetrahedra above 75 GPa.

High-pressure phases of $CaCO_3$ (> 75 GPa) and $MgCO_3$ (> 83 GPa) contain $CO_4^{4-}$ tetrahedra. In $CaCO_3$ above 75 GPa, we see chains of corner-sharing tetrahedra. In $MgCO_3$ above 83 GPa, $CO_4^{4-}$ tetrahedra form $C_3O_9^{6-}$ rings, and above 180 GPa form chains [5,6]. The differences between $CaCO_3$ and $MgCO_3$, come from different sites of Ca and Mg. $Ca^{2+}$ is much larger than $Mg^{2+}$, and requires the anion sublattice to have more open space to fit it.

On the Ca-rich side, our newly predicted phases include stable C*mcm*-$Ca_3CO_5$, *I*4/*mcm*-$Ca_3CO_5$ and metastable $P2_1/m$-$Ca_2CO_4$, all of which contain isolated $CO_4^{4-}$ tetrahedra, as shown in Fig. 5(f)-(h). Our calculations prove that calcium orthocarbonate $Ca_3CO_5$ can be stable at very low pressure (11 GPa), which is much lower than the formation pressure of orthocarbonic acid (314 GPa) [25]. Chemically, $Ca_3CO_5$ can be represented as $CaO \bullet Ca_2CO_4$ with coexistence of both $O^{2-}$ and $CO_4^{4-}$ anions, while metastable $Ca_2CO_4$ is a typical orthocarbonate. Stability of $Ca_3CO_5$ at a surprisingly low pressure of 11 GPa means that the $CO_4^{4-}$ units may also be present at such pressures in carbonate melts. Phonon calculations show that $Ca_3CO_5$ can be quenchable to ambient conditions at low temperatures.

Unlike $Ca_3CO_5$ and $Ca_2CO_4$, with higher $CO_2$ content in $CaC_2O_5$, $CO_4^{4-}$ tetrahedra are connected into 2D-sheets in *Pc*-$CaC_2O_5$ and *C*2-$CaC_2O_5$, and into a 3D-framework in *Fdd*2-$CaC_2O_5$, as shown in Fig. 5(i)-(k). Polymerization of $CO_3^{2-}$ can be described as a transformation from carbonyl (C=O) functional groups to ether bonds (C-O-C), as shown in Fig. 6(a). The charged oxygen atom bonded to carbon is a nucleophilic site, whereas carbon atoms in $CO_2$ molecules are positively charged. This makes an electrophilic reaction possible, with polymerized $CO_3^{2-}$ sharing one electron pair with $CO_2$ upon formation of a polymeric framework $C_2O_5^{2-}$ anion, as shown in Fig. 6(b).

As discussed in previous work [26], the oxygen sharing by $CO_3^{2-}$ and $CO_2$ can be described as oxo-Grotthuss mechanism. Here the formation of $CaC_2O_5$ is its enhanced version. The combination of $CO_3^{2-}$ and $CO_2$ offsets the electrostatic repulsion between $CO_3^{2-}$ anions. This is why the participation of $CO_2$ greatly decreases the polymerization pressure (compared with $CaCO_3$) from 75 to 33 GPa.

## 3. Are $Ca_3CO_5$ and $CaC_2O_5$ possible in the Earth's lower mantle?

By means of the quasi-harmonic approximation (QHA), we first explored thermodynamic stability of all calcium carbonates in the pressure range from 80 GPa to 160 GPa and temperature at 2000 K. We note that, under such pressure and temperature conditions, $CO_2$ is expected to be solid [27]; its Gibbs free energy can thus be accurately computed based on the crystalline structure. As shown in Fig. 7, temperature has a tiny effect on Gibbs free energy of formation of each calcium carbonate and all three calcium carbonates ($Ca_3CO_5$, $CaCO_3$ and $CaC_2O_5$) that are stable at 0 K are still stable at 2000 K, indicating that they will not decompose at the Earth's lower mantle conditions.

Then, we studied the chemical stability of stable $Ca_3CO_5$, $CaC_2O_5$ and $CaCO_3$ and metastable $Ca_2CO_4$ by exploring their possible reactions with compounds $MgSiO_3$ [28,29], $CaSiO_3$ [30,31], $SiO_2$ [32,33] and MgO - we remind that (Mg,Fe)$SiO_3$, $CaSiO_3$ and (Mg,Fe)O are the dominant compounds of the Earth's lower mantle. These reactions are listed below:

$Ca_3CO_5$+3SiO2=3$CaSiO_3$+$CO_2$ (1)

$Ca_3CO_5$+MgO=$MgCO_3$+3CaO (2)

$Ca_3CO_5$+3$MgSiO_3$=3$CaSiO_3$+$MgCO_3$+2MgO (3)

$Ca_3CO_5$+$MgSiO_3$+2$SiO_2$=3$CaSiO_3$+$MgCO_3$ (4)

$Ca_2CO_4$+2$SiO_2$=2$CaSiO_3$+$CO_2$ (5)

$Ca_2CO_4$+MgO=$MgCO_3$+2CaO (6)

$Ca_2CO_4$+2$MgSiO_3$=2$CaSiO_3$+$MgCO_3$+MgO (7)

$Ca_2CO_4$+$MgSiO_3$+$SiO_2$=2$CaSiO_3$+$MgCO_3$ (8)

$CaCO_3$+$SiO_2$=$CaSiO_3$+$CO_2$ (9)

$CaCO_3$+MgO=$MgCO_3$+CaO (10)

$CaCO_3$+$MgSiO_3$=$CaSiO_3$+$MgCO_3$ (11)

$CaC_2O_5$+$SiO_2$=$CaSiO_3$+2$CO_2$ (12)

$CaC_2O_5$+2MgO=CaO+2$MgCO_3$ (13)

$CaC_2O_5$+$MgSiO_3$=$CaSiO_3$+$MgCO_3$+$CO_2$ (14)

$CaC_2O_5$+$MgSiO_3$+MgO=$CaSiO_3$+2$MgCO_3$ (15)

$CaC_2O_5$+MgO=$CaCO_3$+$MgCO_3$ (16)

$CaC_2O_5$+$CaSiO_3$+MgO=2$CaCO_3$+$MgSiO_3$ (17)

For each chemical reaction, we used the most stable structures of each compound at relevant pressures and temperatures. It should be noted that there are some other phases suggested for some major mantle compounds [34], but we would prefer not to take them as references in view of their thermodynamic metastabilities.

Fig. 8 shows the computed Gibbs free energy of each reaction in the pressure range from 80 GPa to 160 GPa and temperature of 2000 K. For the newly predicted $Ca_3CO_5$ and $CaC_2O_5$, we found that they both cannot exist in the Earth's lower mantle. As shown in Fig. 8(a)&(b), $Ca_3CO_5$ will always react with $MgSiO_3$ and $SiO_2$; $CaC_2O_5$ does not react with the main mantle compound $MgSiO_3$, but will react with MgO, and also with a mixture of MgO and $CaSiO_3$. For the well-known $CaCO_3$, we found that $CaCO_3$ will not react with MgO, $SiO_2$, and $MgSiO_3$ at zero temperature and pressure above 90 GPa (see Fig. S3 in Supplementary Materials), in agreement with previous results [5,6]. However, at 2000 K, there is a big change in the behavior of reaction (11) - $CaCO_3$ will always react with $MgSiO_3$ at pressures below 140 GPa, as shown in Fig. 8(c). Fig. 9 shows the phase diagram for reaction (11). It shows that $CaCO_3$ will never exist in the Earth's lower mantle, with large excess of $MgSiO_3$. Therefore, we conclude that throughout the Earth's lower mantle polymorphs of $MgCO_3$ are the main hosts of oxidized carbon.

## Conclusions

In summary, evolutionary crystal structure predictions have been performed for MgO-$CO_2$ and CaO-$CO_2$ systems with the aim of exploring stable magnesium and calcium carbonates at pressures ranging from 0 GPa to 160 GPa. For the MgO-$CO_2$ system, we found that there is only one stable magnesium carbonate $MgCO_3$. For the CaO-$CO_2$ system, in addition to $CaCO_3$, we also discovered two hitherto unknown stable calcium carbonates $Ca_3CO_5$, $CaC_2O_5$ and one near-ground-state compound $Ca_2CO_4$.

$Ca_3CO_5$ can be represented as CaO•$Ca_2CO_4$, and is a calcium orthocarbonate, and is stable at a remarkably low pressure of 11 GPa. This is the lowest-pressure material with $CO_4$-tetrahedra. $CaC_2O_5$ is the product of electrophilic reaction: $CO_3^{2-}$+$CO_2$ and an enhanced version of oxo-Grotthuss mechanism, which greatly decreases the polymerization pressure of $CO_3^{2-}$: 33 GPa, compared to 75 GPa in $CaCO_3$.

We have checked chemical stability of $Ca_3CO_5$ and $CaC_2O_5$ in the Earth's lower mantle environments by investigating possible chemical reactions involving $MgCO_3$, $CO_2$, $MgSiO_3$, $CaSiO_3$, $SiO_2$, CaO, and MgO. Our results indicate that, while chemically very interesting, none of these new carbonates can be present in the Earth's mantle. $Ca_3CO_5$ is so far the lowest pressure stable compound with $CO_4$ tetrahedra and suggests that already at pressures of ~11 GPa, carbonate melts can have a large concentration of $CO_4$ groups.

# References

[1] Oganov A R, Hemley R J, Hazen R M, et al, Reviews in Mineralogy and Geochemistry. 75(1), 47-77(2013)

[2] Suito K, Namba J, Horikawa T, Taniguchi Y, Sakurai N, Kobayashi M, Onodera A, Shimomura O, Kikegawa T, American Mineralogist. 86, 997–1002( 2001)

[3] Oganov A R, Glass C W, Ono S, Earth and Planetary Science Letters . 241, 95-103( 2006)

[4] Fiquet G, Guyot F, Kunz M, Matas J, Andrault D, and Hanfland M, American Mineralogist. 87, 1261 (2002).

[5]Oganov A R, Ono S, Ma Y, Glass C W, Garcia A, Earth and Planetary Science Letters. 273, 38-47(2008)

[6] Pickard C J, Needs R J, Physical Review B. 91, 104101(2015)

[7] Zhu Q, Jung D Y, Oganov A R, et al., Nature chemistry. 5(1): 61-65(2013)

[8] Zhang W, Oganov A R, Goncharov A F, et al., Science. 342(6165): 1502-1505(2013)

[9] Dong X, Oganov A R, Goncharov A F, et al., Nature Chemistry. 9(5): 440-445( 2017)

[10] Oganov A R, Glass C W, The Journal of Chemical Physics. 124, 244704( 2006)

[11] Oganov A R, Lyakhov A O, Valle, M, Accounts of Chemical Research. 44, 227-237( 2011)

[12] Lyakhov A O, Oganov A R, Stokes H T, Zhu Q, Computer Physics Communications. 184, 1172-1182(2013)

[13] Glass C W, Oganov A R, Hansen N, Computer Physics Communications. 175, 713-720(2006)

[14] Kresse G, Furthmüller J, Physical Review B. 54(16), 11169-11186(1996)

[15] Kohn W, Sham LJ, Physical Review. 140, A1133(1965)

[16] Perdew J P, Burke K, Ernzerhof M, Physical Review Letters. 77(18), 3865-3868( 1996)

[17] Blöchl P E, Physical Review B. 50(24), 17953-17979(1994)

[18] Jeanloz R, Ahrens T J, Mao H K, et al, Science, 206(4420), 829(1979)

[19] Duffy T S, Hemley R J, Mao Hk, Physical Review Letters. 74, 1371-1374(1995)

[20] Bonev S A, Gygi F, Ogitsu T, Galli G, Physical Review Letters. 91, 065501(2003)

[21] Dean Smith, Keith V. Lawler, Austin W. Daykin, et al, Physical Review Materials 2, 013605

[22] Ono S, Kikegawa T, Ohishi Y, Tsuchiya J, American Mineralogist. 90, 667–671(2005)

[23] Merlini, M., M. Hanfland, and W. A. Crichton, Earth & Planetary Science Letters. 333-334, 265-271(2012).

[24] Momma K, Izumi F, Journal of Applied Crystallography. 44(6), 1272-1276(2011)

[25] Saleh G, Oganov A R, arXiv preprint arXiv. 1603.02425(2016)

[26] Corradini D, Coudert F X, Vuilleumier R, Nature chemistry. 8(5), 454-460(2016)

[27] Boates B, and Teweldeberhan A M, and Bonev S A, Proceedings of the National Academy of Sciences. 109(37), 14808-14812(2012)

[28]Tsuchiya T, Tsuchiya J, Umemoto K, Wentzcovitch R M, Earth and Planetary Science Letters. 224, 241-248(2004)

[29] Oganov A R, Ono S, Nature. 430(6998), 445-448(2004)

[30] Jung D Y, Oganov A R, Physics and Chemistry of Minerals. 32, 146-153(2005)

[31] Serghiou G C, Hammack W S, The Journal of Chemical Physics. 98, 9830-9834(1993)

[32] Marto&Ncaron ák R, Donadio D, Oganov A R, et al, Nature Materials. 5(8),623-626(2006)

[33] Oganov A R, Gillan M J, Price G D, Physical Review B. 71, 064104(2005)

[34] Magyari-Kope B, Vitos L, Grimvall G, Johansson B, and Kollar J, Physcial Review B 65, 193107(2002)

# Acknowledgment

This work was partly supported by National Science Foundation (grant EAR-1723160) and Foreign Talents Introduction and Academic Exchange Program of China (No. B08040). The authors acknowledge the High Performance Computing Center of NPU for the allocation of computing time on their machines. X. D. also acknowledges access to Tianhe II supercomputer in Guangzhou.

**Figure legend:**

Fig. 1 Thermodynamic convex hulls for $MgO$-$CO_2$ and $CaO$-$CO_2$ systems at zero temperature and high pressure (with zero-point energy correction).

Fig. 2 Phonon dispersion curves of the newly predicted calcium carbonates at high pressures and zero temperature.

Fig. 3 Pressure-composition phase diagrams of $MgO$-$CO_2$ and $CaO$-$CO_2$ system at zero temperature.

Fig. 4 Equations of state of all stable calcium carbonates at zero temperature.

Fig. 5 Crystal structures of predicted stable and metastable calcium carbonates.

Fig. 6 Mechanism of (a) the polymerization of $CO_3^{2-}$ and (b) the formation of $CaC_2O_5$.

Fig. 7 Thermodynamic convex hulls for the $CaO$-$CO_2$ system at 2000 K and various pressures.

Fig. 8 Gibbs free energies of mantle-relevant reactions as a function of pressure (at 2000 K).

Fig. 9 Relative stability of the $MgCO_3$ + $CaSiO_3$ assemblage versus $CaCO_3$ + $MgSiO_3$.

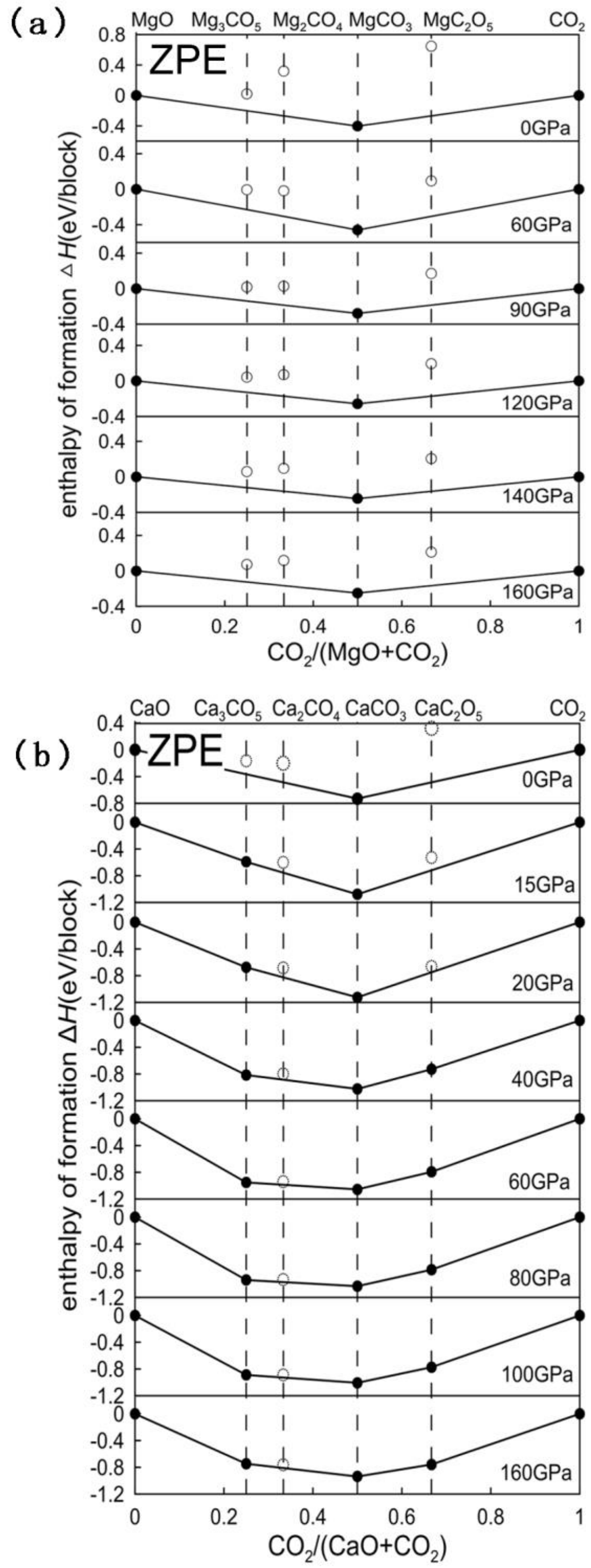

Fig. 1 Thermodynamic convex hulls for $MgO$-$CO_2$ and $CaO$-$CO_2$ systems at zero temperature and high pressure (with zero-point energy correction). Filled circles denote stable structures and open circles denote metastable structures. Enthalpies of formation from oxides are normalized to one oxide unit.

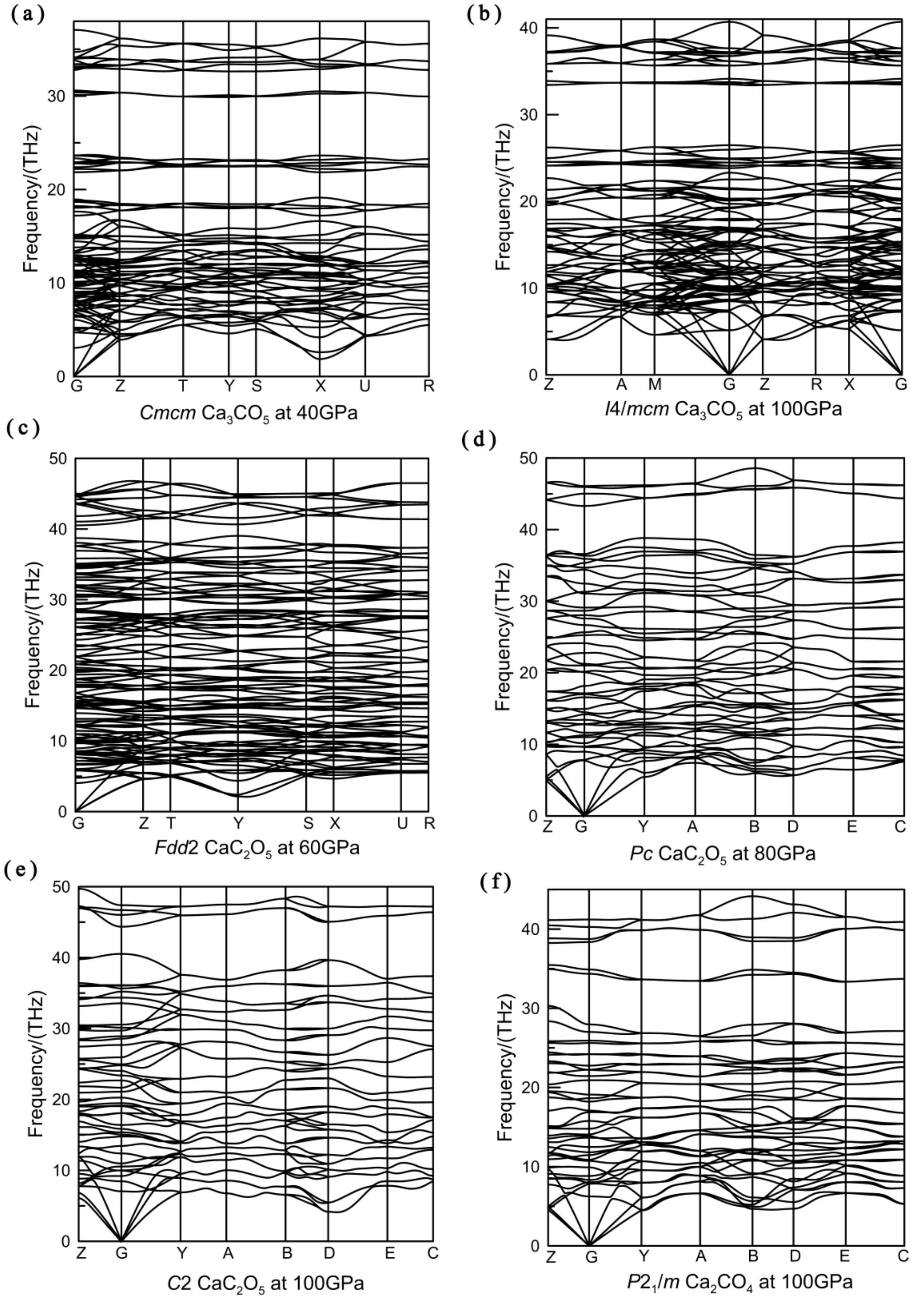

Fig. 2 Phonon dispersion curves of the newly predicted calcium carbonates at high pressures and zero temperature.

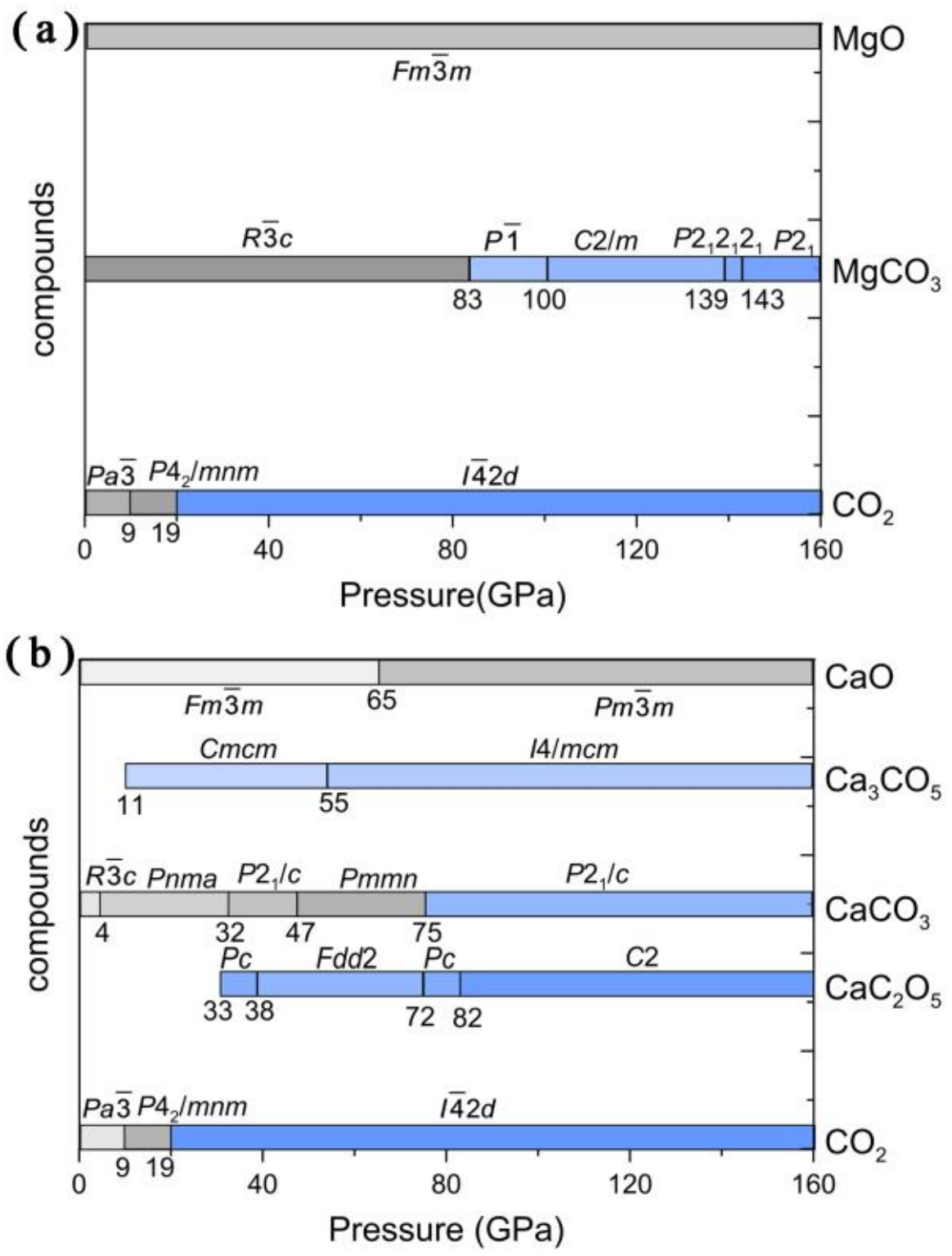

Fig. 3 Pressure-composition phase diagrams of MgO-$CO_2$ and CaO-$CO_2$ system at zero temperature.

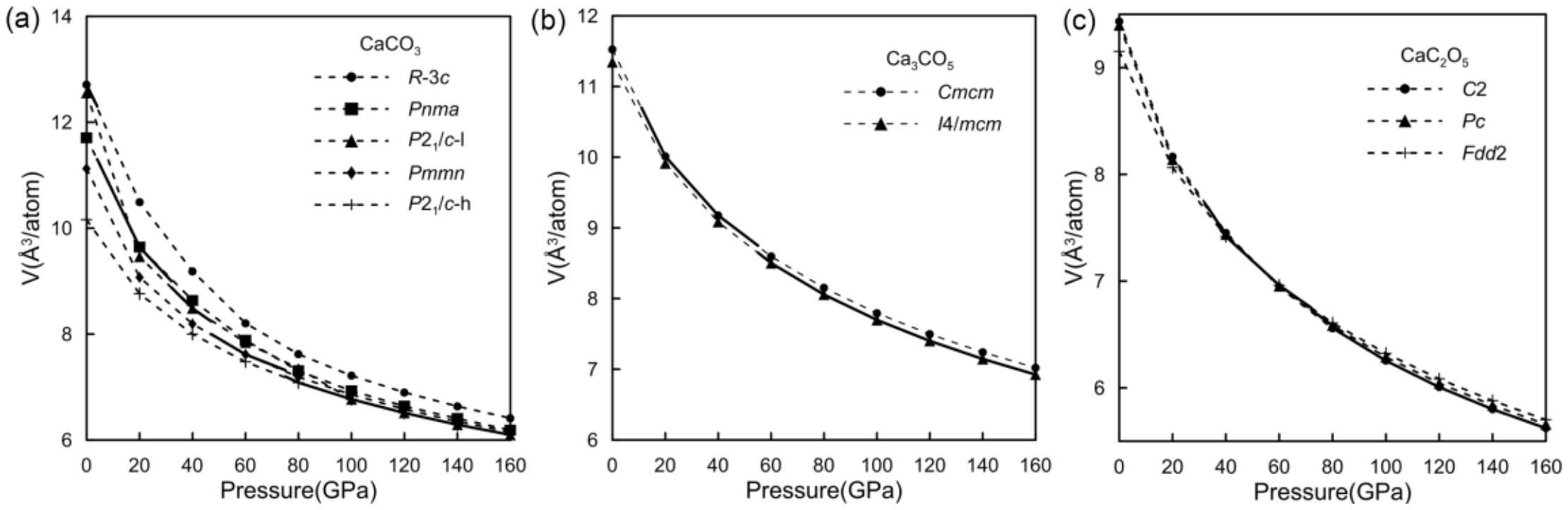

Fig. 4 Equations of state of all stable calcium carbonates at zero temperature. Solid lines denote equations of state of each phase in its region of stability.

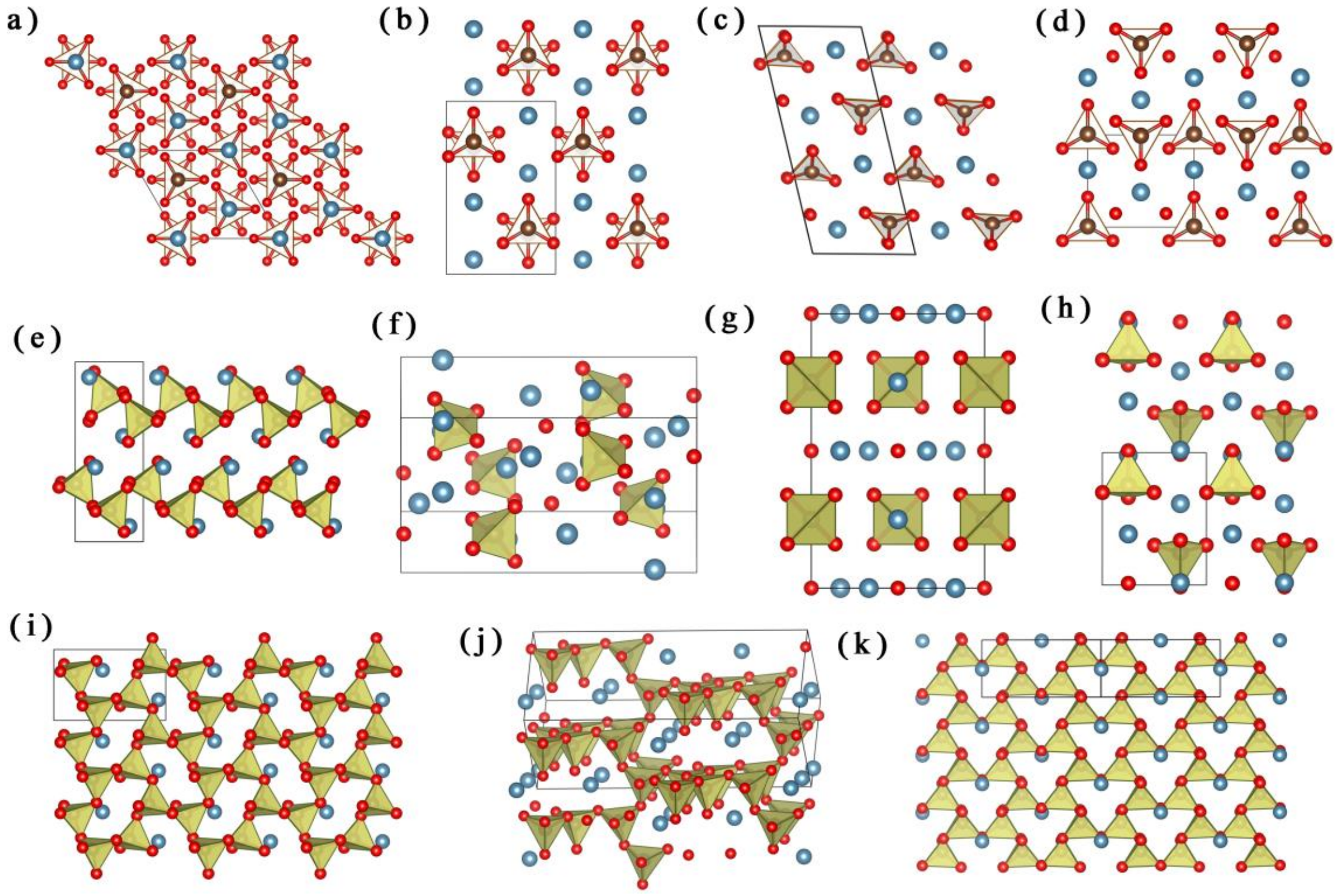

Fig. 5 Crystal structures of predicted stable and metastable calcium carbonates. (a) *R*-3*c* (calcite); (b) *Pnma* (aragonite); (c) $P2_1/c$-l; (d) *Pmmn* (post-aragonite); (e) $P2_1/c$-h; (f) *Cmcm* $Ca_3CO_5$; (g) *I*4/*mcm* $Ca_3CO_5$; (h) $P2_1/m$ $Ca_2CO_4$; (i) *Pc* $CaC_2O_5$; (j) *Fdd*2 $CaC_2O_5$; (k) *C*2 $CaC_2O_5$

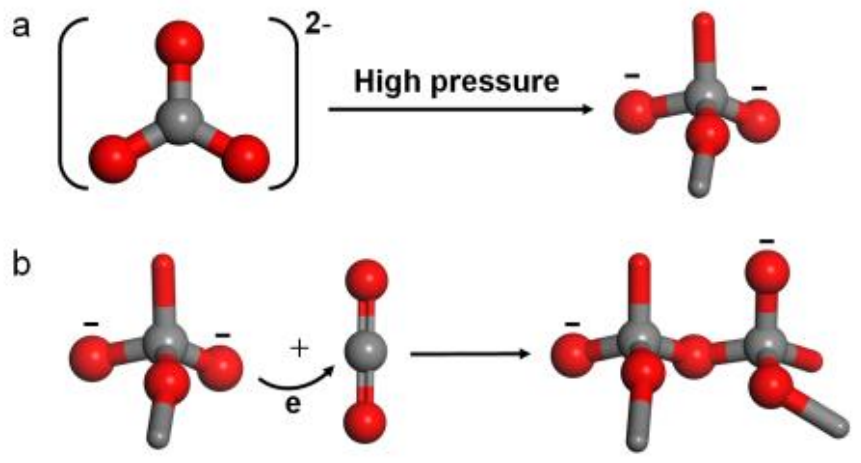

Fig. 6 Mechanism of (a) the polymerization of $CO_3^{2-}$ and (b) the formation of $CaC_2O_5$.

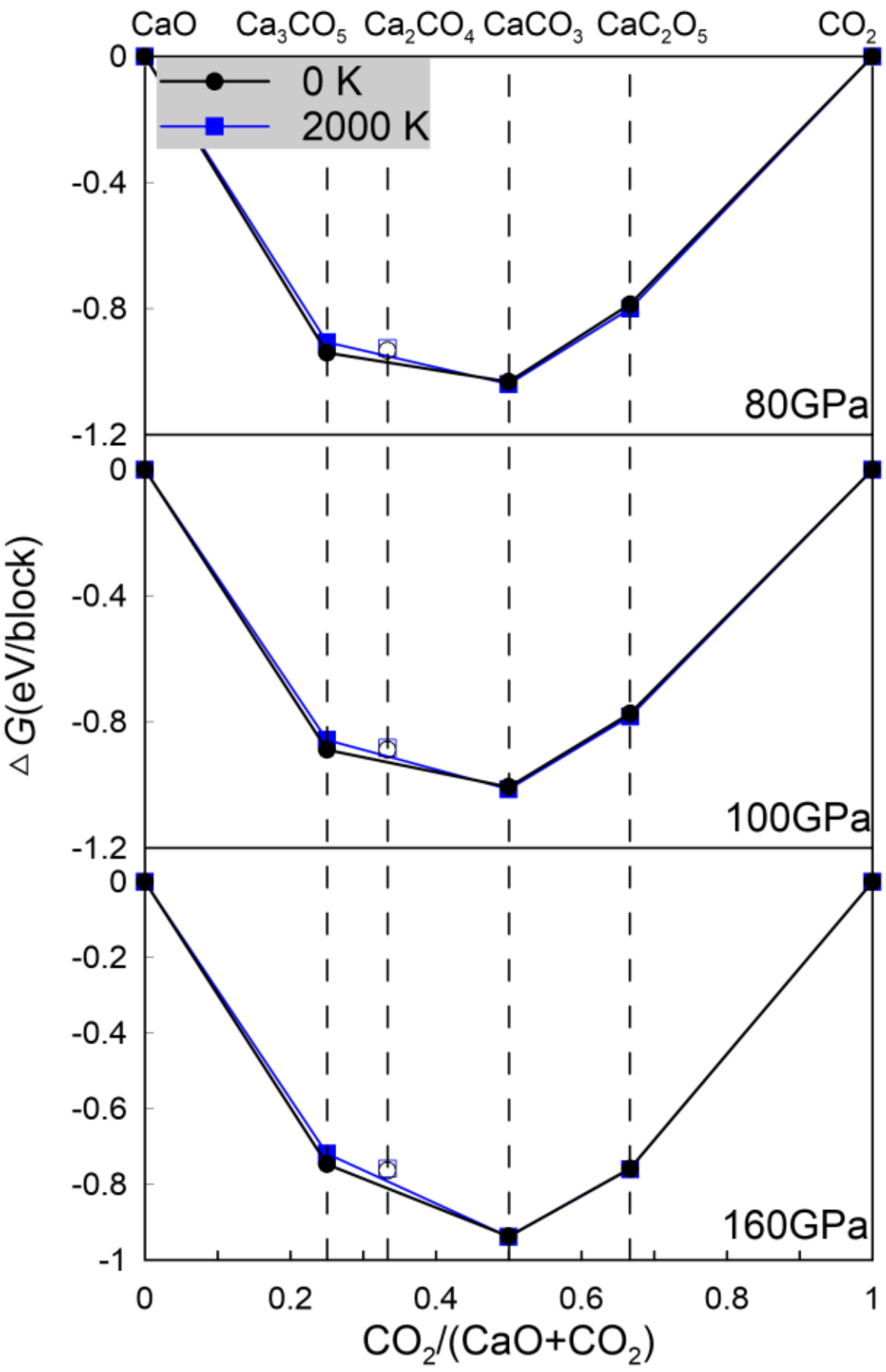

Fig. 7 Thermodynamic convex hulls for the CaO-$CO_2$ system at 2000 K and various pressures. Filled symbols denote stable structures, open symbols - metastable structures. Gibbs free energies of formation from oxides are normalized to one oxide unit.

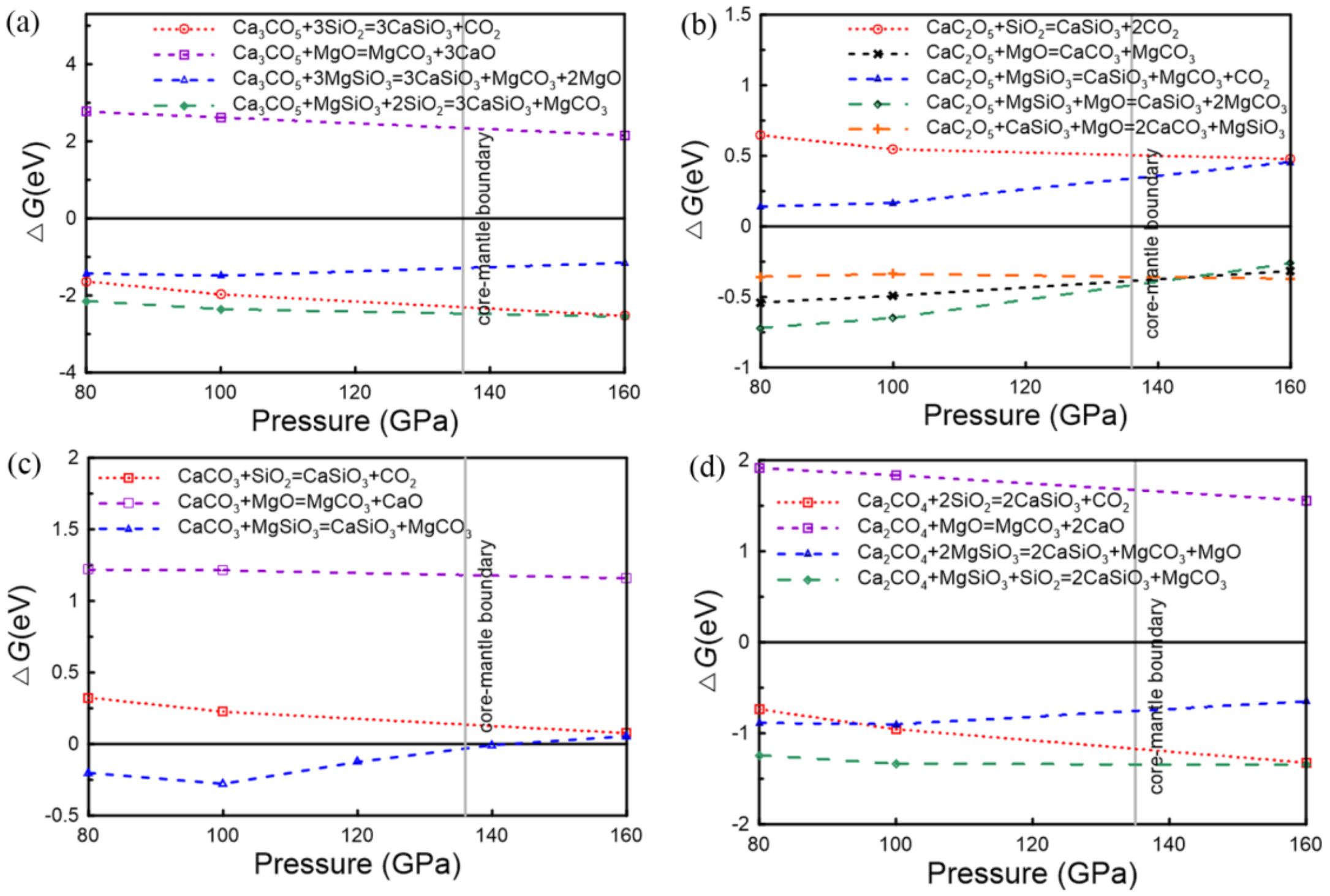

Fig. 8 Gibbs free energies of mantle-relevant reactions as a function of pressure (at 2000 K).

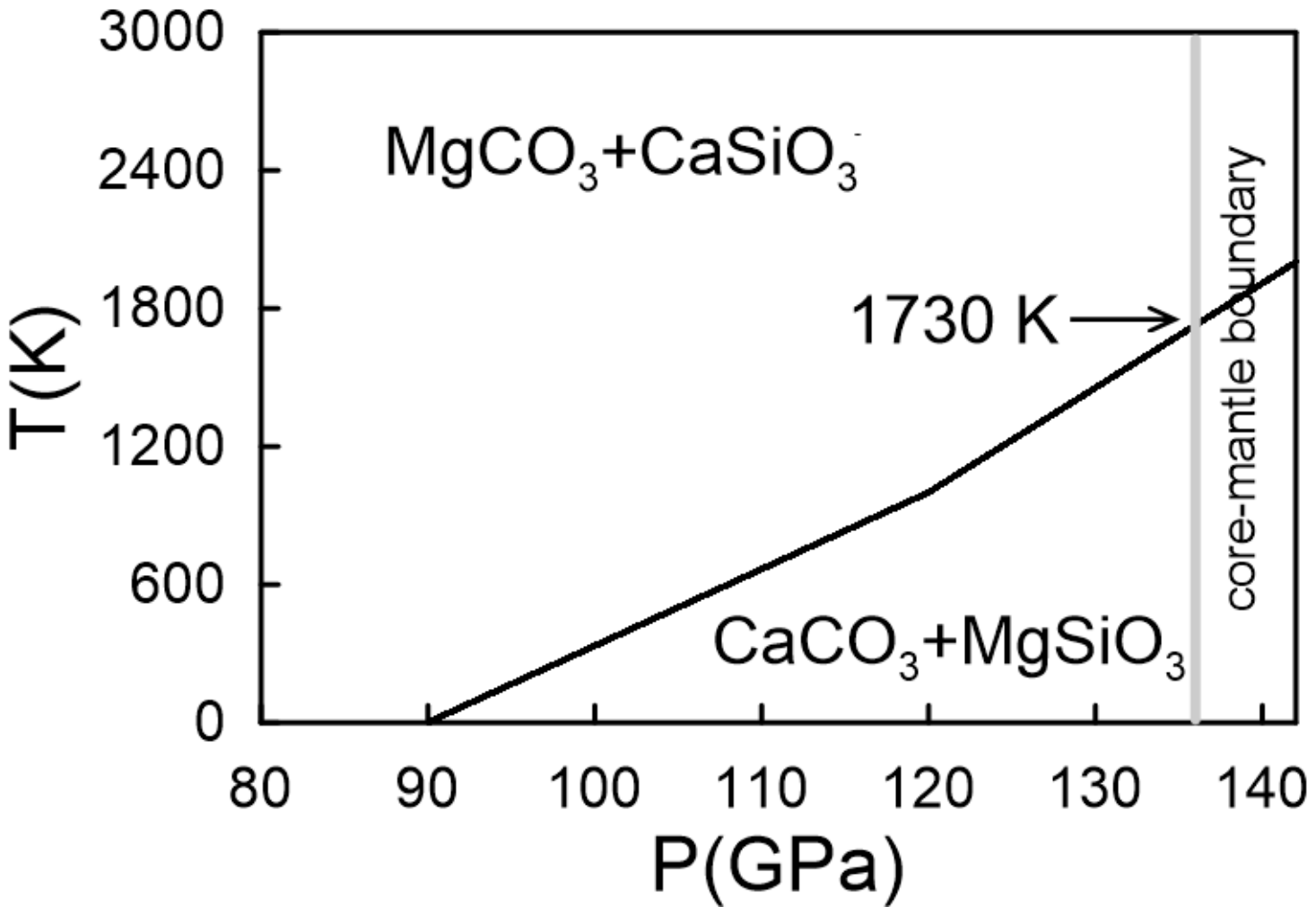

Fig. 9 Relative stability of the $MgCO_3$ + $CaSiO_3$ assemblage versus $CaCO_3$ + $MgSiO_3$.